\documentclass[aps,prd,twocolumn,groupedaddress,nofootinbib]{revtex4-1}

\usepackage{dsfont} \usepackage{amssymb} \usepackage{graphicx} \usepackage{amsmath} \usepackage{url} \usepackage{bm} \usepackage{epstopdf}
\usepackage{color} \usepackage{enumitem} \usepackage[colorlinks=true,breaklinks]{hyperref}
\usepackage{dsfont} \usepackage{mathrsfs} \usepackage[utf8]{inputenc}

\newcommand{\di}{\partial}  \newcommand{\eff}{\text{eff}} \newcommand{\Pl}{\text{Pl}}
 \renewcommand{\L}{\mathcal{L}}
\newcommand{\D}{\mathcal{D}}

\def\be {\begin{equation}}

\def\ee  {\end{equation}}

\def\bea {\begin{eqnarray}}

\def\eea {\end{eqnarray}}

\begin{document}

\title{Gravitational wave defocussing in quadratic gravity}

\author{Jack Gegenberg} \email{geg@unb.ca} \affiliation{Department of Mathematics and Statistics, University of New Brunswick, Fredericton, NB,
Canada E3B 5A3}

\author{Sanjeev S.\ Seahra} \email{sseahra@unb.ca} \affiliation{Department of Mathematics and Statistics, University of New Brunswick,
Fredericton, NB, Canada E3B 5A3}

\date{\today}

\begin{abstract}

We demonstrate that Huygens' principle for gravitational waves fails in quadratic gravity models that exhibit conformal symmetry at high
energies.  This results in the blurring of gravitational wave signals over finite timescales related to the energy scale of new physics
$M_{\star}$.  Furthermore, on very small scales the gravitational wave Green's function reduces to that of the wave equation in two dimensions.
In principle, $M_{\star}$ could be constrained directly from gravitational wave observations.

\end{abstract}

\maketitle

\section{Introduction}\label{sec:introduction}

The recent direct detection of gravitational waves by LIGO \cite{Abbott:2016blz} provides us with new ways of testing Einstein's general
relativity (GR) \cite{Yunes:2013dva,*TheLIGOScientific:2016src,*Yunes:2016jcc,*TheLIGOScientific:2016pea}.  We are now able to directly measure
the time-dependent, weak fluctuations of spacetime geometry first predicted over 100 years ago.  A key prediction of GR is that, in many
respects, these perturbations behave exactly like waves of electromagnetic radiation:   Small amplitude gravitational waves propagate in
sharply-defined spherical wavefronts traveling at a fixed speed.  That is, gravitational waves obey the same Huygens' principle that underpins
our understanding of sound and light waves in three spatial dimensions.

A natural and interesting question is: does Huygens' principle hold in theories of gravitation other than GR?  Many such alternative theories
\cite{Clifton:2011jh} have been proposed, and they are often motivated by the failure of GR coupled to ordinary matter to explain observations of
the universe at the largest scales.  Others follow from the observation that naive attempts to quantize gravity are not renormalizable.
Regardless of motivation, viable modified gravity theories tend to leave the predictions of GR unchanged on solar system scales, while
significantly altering them over much smaller or larger distances.  (See \cite{Berti:2015itd} for a recent review on observational tests of
modified theories of gravity.)

One procedure for constructing alternate gravitational models involves adding functions of curvature scalars to the Einstein-Hilbert Lagrangian.
These terms can be selected to accomplish a specific goals, as in many $f(R)$ models \cite{Sotiriou:2008rp}, or they can be derived in an
approximation to a more fundamental framework, as in string theory.  In this work, we consider actions of the form
\begin{equation}\label{eq:action 1} 	S =  \frac{M_\Pl^{2}}{2} \int d^{4}x \sqrt{-g} \, \left(R - 2\Lambda - \frac{2}{M_{\star}^{2}} C^2
\right)+ S_\text{m}. \end{equation} Here, $C^{2} =
C^{\alpha\beta\gamma\delta} C_{\alpha\beta\gamma\delta}$ is the square of the Weyl tensor, $M_{\star}$ is a fixed mass that sets the energy scale
of new physics, $\Lambda = 3H^{2}$ is the cosmological constant ($\Lambda>0$ for de Sitter space), $S_\text{m}$ is the matter action, and
$M_\text{Pl}^{2} = 1/8\pi G$ is the reduced Planck mass.  This action belongs to the class of quadratic gravity theories
\cite{Stelle:1976gc,*Stelle:1977ry,Boulware:1983td,*David:1984uv,*Horowitz:1984wv,*Deser:2003up,*Deser:2007vs,*tHooft:2011aa,*Maldacena:2011mk,*Capozziello:2015nga,*Hinterbichler:2015soa,*Gegenberg:2015gma,Alvarez-Gaume:2015rwa};
i.e., models whose gravitational actions involving quadratic functions of curvature scalars.  Any locally de Sitter solution of general
relativity is a vacuum ($S_\text{m}=0$) solution of (\ref{eq:action 1}), including the family of Kerr black holes with cosmological constant.

We note that if we were to include an additional $R^{2}$ term in (\ref{eq:action 1}), we would have the most general form of quadratic
gravity action.  Such a term basically represents an $f(R)$ contribution to the theory.  For the linear perturbations that we consider below,
such a contribution is fairly well studied \cite{Berry:2011pb}.  We therefore omit this term for simplicity.\footnote{If we included the $R^{2}$ term in the action, there would be an additional spin-0 degree of freedom in the theory \cite{Berry:2011pb}.  Since modes of different helicity are decoupled in linear theory, metric perturbations from the $R^{2}$ term would just be \emph{added} to the spin-2 fluctuations presented in this paper.}

The spectrum of gravitational excitations about flat space in quadratic gravity is well understood \cite{Stelle:1976gc,*Stelle:1977ry}.  In
addition to the massless graviton of GR, the Weyl-squared term in the action induces a massive spin-2 field.  When the action is expanded to
second order in fluctuations, the sign of the massive graviton kinetic term is opposite to that of the massless graviton, indicating the presence
of a quantum ghost instability.  Though we will be exclusively concerned with classical physics in this work, this ghost mode implies that we
should view (\ref{eq:action 1}) as an effective, as opposed to a fundamental, action.  (A recent discussion of the ghost phenomenon in quadratic gravity can be found in \cite{Alvarez-Gaume:2015rwa}, for example.)  We do not speculate on the nature of the fundamental
theory giving rise to (\ref{eq:action 1}), other than assuming that it is ghost-free and (\ref{eq:action 1}) is valid for scales $\gtrsim
1/M_{\star}$, and may indeed be applicable over a range of shorter distances.\footnote{A classical theory that involves higher order curvature terms yet reduces to (\ref{eq:action 1}) on scales $\gtrsim \alpha_\text{m} M_{\star}^{-1}$ (with $\alpha_\text{m} \ll 1$) can be constructed as follows:  Assume a gravitational Lagrangian $\mathcal{L} = R - 2\Lambda -2 M_{\star}^{-2} C^{2} + \alpha_{3}M_{\star}^{-4}  \mathcal{L}_{3} + \alpha_{4}M_{\star}^{-6}  \mathcal{L}_{4} \cdots$ where $\mathcal{L}_{3}$ is a term involving cubic curvature terms, $\mathcal{L}_{4}$ is a term involving quartic curvature terms, etc., and the $\alpha_{n}$ are dimensionless parameters.  Then, assume a hierarchy $\text{max}\{|\alpha_{3}|, |\alpha_{4}|, \ldots\} = \alpha_\text{m} \ll 1$.  It then follows that the dominant part of the action on scales $\gtrsim \alpha_\text{m} M_{\star}^{-1}$ is exactly (\ref{eq:action 1}).}

We can see that the $C^{2}$ term in (\ref{eq:action 1}) is negligible on large scales in the context of linear theory by a heuristic argument that will be confirmed more rigorously below:  Linear metric fluctuations about symmetric backgrounds, such as Minkowski or de Sitter, may be expanded in terms of Fourier modes $\propto e^{i k_{\alpha} x^{\alpha} }$.  Since the Weyl term in (\ref{eq:action 1}) involves the square of second order derivatives of the metric, it will be negligible for long wavelength modes ($\max\{|k_{0}|,|\mathbf{k}|\} \ll M_{\star}$) and the action will reduce to the familiar Einstein-Hilbert form.  Conversely, for gravitational modes with short wavelengths ($\min\{|k_{0}|,|\mathbf{k}|\} \gg M_{\star}$) we expect the $C^{2}$ term to dominate: 
\begin{equation}\label{eq:Weyl action} 	S \approx  -g_\eff^{2} \int d^{4}x \sqrt{-g} \,
C^2 + S_\text{m}, \quad g_{\eff} = \frac{M_\Pl}{M_{\star}}, \end{equation} which yields the conformally invariant gravitational action originally
written down by Weyl in 1918.

Such conformal theories are attractive because they involve no fixed length scales; that is, only angles are physically meaningful.  Hence, all
coupling constants are necessarily dimensionless, suggesting that these gravitational models are perturbatively renormalizable, unlike ordinary
GR.  However, if the universe really does possess an exact conformal symmetry, it must be spontaneously broken such that one recovers an
effective Einstein-Hilbert action on astrophysical scales \cite{Hooft:2014daa}.  Equation (\ref{eq:action 1}) is a plausible ansatz for the
action of theory with a broken conformal symmetry that behaves like GR in the infrared.

Now, since the Newtonian limit of GR has been verified in the laboratory on scales $\gtrsim 10^{-8}\,\text{m}$ \cite{Will:2014kxa}, this
immediately suggests an upper bound on the Compton wavelength of the exotic physics scale: $M_{\star}^{-1} \lesssim 10^{-8}\,\text{m}$.   Enforcing this bound implies that the influence of the $C^{2}$ term on solar system tests or the phase evolution of binary inspirals will be negligibly small.

We can now attempt to anticipate the behaviour of classical gravitational waves governed by (\ref{eq:action 1}).  As in GR, the details of wave
propagation will be dictated by the Green's function $G(x,x')$ associated with the relevant equation of motion.   Due to local Lorentz
invariance, the Green's function can only be a function of the geodesic distance $\ell(x,x')$ between the field point $x$ and the source point
$x'$.  When $|\ell| \gg 1/M_{\star}$, we should recover the Green's function of ordinary GR with a cosmological constant.  However, the behaviour
of $G$ for $|\ell| \ll 1/M_{\star}$ will be dictated by the conformally invariant part of (\ref{eq:action 1}); i.e., by the action (\ref{eq:Weyl
action}).  The crucial observation is as follows:  The nullcone connecting $x$ and $x'$ is defined by $\ell(x,x') = 0$, which means that
structure of the Green's function for null and nearly-null separations is entirely determined by the Weyl-squared part of the action.

But Huygens' principle is essentially the statement that the Green's function has singular support on the future nullcone of the source point
$x'$.  Therefore, we have the strong suspicion that the principle
will modified in the theory of gravity (\ref{eq:action 1}).  In the following section, we demonstrate that this intuition is correct: Huygens'
principle does is not valid for gravitational waves governed by
(\ref{eq:action 1}) because their propagator is non-singular and has support for timelike separations of the field and source points.  The
$\delta$-function gravitational wave nullcone of GR is smeared out
over a distance $\sim 1/M_{\star}$, implying an intrinsic blurriness of the signals detected by experiments such as LIGO.

\section{Gravitational wave Green's function}\label{sec:Green's}

Variation of the action (\ref{eq:action 1}) with respect to the metric yields the equation of motion \begin{equation}\label{eq:field eqn} 	
G_{\alpha\beta} + \Lambda g_{\alpha\beta} + 2 M_{\star}^{-2}
B_{\alpha\beta} = {M_\Pl^{-2}} T_{\alpha\beta}, \end{equation} where $G_{\alpha\beta}$ and $T_{\alpha\beta}$ are the Einstein and stress-energy
tensors as usual, while $B_{\alpha\beta}$ is the Bach
tensor defined by: \begin{subequations} \begin{align} B_{\mu\nu} & = -\nabla^{\alpha} \nabla_{\alpha} S_{\mu\nu} + \nabla^{\alpha} \nabla_{\mu}
S_{\alpha\nu} + C_{\mu\alpha\nu\beta} S^{\alpha\beta}, \\
S_{\alpha\beta} & = \tfrac{1}{2}(R_{\alpha\beta} - \tfrac{1}{6} R g_{\alpha\beta}). \end{align} \end{subequations} Note that the Bach tensor
involves fourth order derivatives of the metric, hence the action
(\ref{eq:action 1}) represents a higher derivative theory of gravity.

The Bach tensor transforms trivially under conformal transformations, implying that $B_{\alpha\beta} = 0$ for conformally flat spacetimes.
Assuming that $\Lambda > 0$, the vacuum solution of (\ref{eq:field
eqn}) with $T_{\alpha\beta} = 0$ is de Sitter space, which we take as the background geometry for perturbations.  We work in the conformal time
coordinate chart, given by \begin{equation}\label{eq:conformal
line element} 	ds^{2} = (H\eta)^{-2} (-d\eta^{2}+d\mathbf{x}^{2}). \end{equation} An important quantity in de Sitter space for the calculation
of Green's functions is the geodesic distance $\ell(x,x')$ between
spacetime points $x = (\eta,\mathbf{x})$ and $x'=(\eta',\mathbf{x}')$.  This is given implicitly by the formula \cite{Tsamis:1992xa}:
\begin{equation}\label{eq:geodesic distance} 	\sin^{2} \left[ \tfrac{1}{2} H
\ell(x,x') \right] = \tfrac{1}{4} (\eta\eta')^{-1} [-(\Delta\eta)^{2} + r^{2}], \end{equation} where $\Delta\eta = \eta - \eta'$ and $r =
|\mathbf{x}-\mathbf{x}'|$.  We note that in the limit that the geodesic distance is
much less than the de Sitter curvature scale $H|\ell| \ll 1$, in which case we approach the flat spacetime limit, we have \begin{equation} 	
\ell^{2}(x,x') \approx -(\Delta\eta)^{2} + r^{2}, \quad
H\eta\approx 1, \quad H\eta' \approx 1. \end{equation}

We consider gravitational wave fluctuations about such a de Sitter background.  We will \emph{a priori} assume that the perturbations are transverse and traceless:
\begin{equation}\label{eq:gauge choice} 	
\delta g_{\alpha\beta} = h_{\alpha\beta}, \quad
g^{\alpha\beta}h_{\alpha\beta} = 0 = \nabla^{\alpha} h_{\alpha\beta}.
\end{equation}
Inserting (\ref{eq:gauge choice}) into the field equations and expanding to linear order in $h_{\alpha\beta}$, one finds that the metric
perturbation obeys a fourth order wave equation \begin{equation} 	(2M_{\star}^{2})^{-1} ( D^{2} - 2H^{2})(D^{2} - 4H^{2} -
M_{\star}^{2})h_{\alpha\beta} = 0, \end{equation} where $D^{2} = \nabla_{\alpha}
\nabla^{\alpha}$ is the covariant wave operator in de Sitter space.  We further assume that $h_{\alpha\beta}$ is purely spatial such that the
perturbed line element is 
\begin{equation}\label{eq:perturbed line
element} 	ds^{2} = (H\eta)^{-2} \left[ -d\eta^{2}+ (\delta_{ij} + H\eta\, \psi_{ij}) dx^{i} dx^{j} \right], \end{equation} where $i,j=1,2,3$.
It is straightforward to confirm that each component of the spatial tensor
$\psi_{ij}$ satisfies a fourth-order partial differential equation: \begin{equation}\label{eq:explicit EOM} 	\frac{H^{3}\eta}{2M_{\star}^{2}}
\left( \Box - \frac{M_{\star}^{2}}{H^{2}\eta^{2}} \right)
\eta^{2}  \left( \Box + \frac{2}{\eta^{2}} \right) \psi_{ij} = 0. \end{equation} Here, $\Box = -\di_{\eta}^{2} +\mathbf{\nabla}^{2} = -\di_{\eta}^{2} + \di_{x}^{2}+\di_{y}^{2}+\di_{z}^{2}$ is the ordinary d'Alembertian operator of flat space.

We note that our assumptions about the metric perturbations mean that we are considering gravitational waves with the same polarizations as
found in GR, which is certainly not the most general pertubative ansatz.  Of course, it would be interesting to consider other gravitational wave polarizations that may be expected due to the presence of an effective massive graviton \cite{Stelle:1976gc,*Stelle:1977ry}; however, for simplicity, we defer that discussion for future work.

It is useful to re-write (\ref{eq:explicit EOM}) in terms of differential operators:
\begin{equation} 	
\L_{x} \psi_{ij} (x) = 0, \quad \L_{x} = \D_{2,x} \L_{2,x} \D_{1,x} \L_{1,x}, \end{equation} where
\begin{align} \nonumber 	\L_{1,x} & = \Box + 2/\eta^{2}, & \D_{1,x} & = \eta^{2}, \\ 	\L_{2,x} & = \Box - \zeta/\eta^{2}, & \D_{2,x} & =
H\eta/2\zeta, \end{align} and $\zeta = M_{\star}^{2}/H^{2}$.  The
subscript $x$ notation on the operators is meant to convey that they act with respect to the $x = (\eta,\mathbf{x})$ spacetime point.

Now, we note that the inverses of $\D_{1,x}$ and $\D_{2,x}$ are trivial:
\begin{equation} 	\D_{1,x}^{-1} = 1/\eta^{2}, \quad \D_{2,x}^{-1} = 2\zeta/H\eta,
\end{equation}
and we have the following relationships between operators \begin{subequations}\label{eq:identity} \begin{align}\label{eq:identity 1} 	\L_{1,x}
- \L_{2,x} & = (2+\zeta) \D_{1,x}^{-1}, \\ \label{eq:identity 2} \L_{1,x} \D_{1,x} \L_{2,x} & = \L_{2,x}
\D_{1,x} \L_{1,x}. \end{align} \end{subequations} We now show how the identities (\ref{eq:identity}) allow us to calculate the Green's function
$G$ of $\L$ given knowledge of the Green's functions $G_{1}$ and
$G_{2}$ of $\L_{1}$ and $\L_{2}$, respectively.  The respective Green's functions satisfy \begin{subequations} \begin{align} 	\label{eq:G def
1} \L_{x} G(x,x') & = \delta^{(4)}(x-x'), \\ 	\label{eq:G def 2}
\L_{1,x} G_{1}(x,x') & = \delta^{(4)}(x-x'), \\ 	\label{eq:G def 3} \L_{2,x} G_{2}(x,x') & = \delta^{(4)}(x-x'). \end{align}
\end{subequations} We claim that \begin{equation}\label{eq:Greens 1} 	G(x,x') =
\frac{2M_{\star}^{2} [G_{2}(x,x') - G_{1}(x,x') ]}{H\eta' \,(M_{\star}^{2}+2H^{2})}; \end{equation} or, re-written in a more compact form:
\begin{equation}\label{eq:Greens 2} 	G_{xx'} = (2+\zeta)^{-1}
(G_{2,xx'} - G_{1,xx'}) \D_{2,x'}^{-1}, \end{equation} where $G_{xx'} = G(x,x')$, etc.  To establish that (\ref{eq:Greens 1}) is indeed a
solution of (\ref{eq:G def 1}), we operate $\L$ on (\ref{eq:Greens 2}):
\begin{eqnarray} 	\L_{x} G_{xx'} & = &  \L_{x} (2+\zeta)^{-1} (G_{2,xx'} -G_{1,xx'}) \D_{2,x'}^{-1}  \nonumber 	\\ & = &(2+\zeta)^{-1}
\D_{2,x} \D_{2,x'}^{-1} (\L_{1,x} \D_{1,x} \L_{2,x} G_{2,xx'} \nonumber
\\ & & - \L_{2,x} \D_{1,x} \L_{1,x} G_{1,xx'})  \nonumber 	\\ & = & (2+\zeta)^{-1} \D_{2,x}  \D_{2,x'}^{-1} (\L_{1,x} - \L_{2,x}) \D_{1,x}
\delta^{(4)}_{xx'} \nonumber 	\\ & = & \delta^{(4)}_{xx'},
\end{eqnarray} where we have used (\ref{eq:identity 2}) to go from the first to the second line, (\ref{eq:identity 1}) to go from the third to
the fourth line, and the shorthand notation $\delta^{(4)}_{xx'}
= \delta^{(4)}(x-x')$.  Hence, our claim (\ref{eq:Greens 1}) is confirmed.

At first glance, (\ref{eq:Greens 1}) may be surprising since it states that the Green's function of the fourth order differential operator $\L$
is given by a superposition of the Green's functions of two second
order operators $\L_{1}$ and $\L_{2}$.  However, we note that $G_{1}$ and $G_{2}$ are the Green's functions for massless and massive gravitons in
de Sitter space, respectively, which are both expected fields in
the spectrum of quadratic gravity.  Furthermore, the relative sign between $G_{1}$ and $G_{2}$ is reminiscent of the sign difference in the
kinetic terms of these modes that leads to the ghost mode in the
quantum version of the theory.   We will soon see that this relative sign means the singular parts of $G_{1}$ and $G_{2}$ cancel out in
(\ref{eq:Greens 1}), making $G$ non-singular.

The Green's function is not completely specified by (\ref{eq:Greens 1}) until we select boundary conditions.  Causality dictates that the
relevant Green's function for gravitational wave propagation satisfies
retarded boundary conditions; that is, $G(x.x')$ vanishes if $x$ is not in the causal future of $x'$.  We can satisfy this by selecting $G_{1}$
and $G_{2}$ to be the retarded Green's functions of $\L_{1}$ and
$\L_{2}$, respectively.  Fortunately, it is relatively easy to find $G_{1}$ and $G_{2}$ by either inverting the second order operators using
standard techniques \cite{Poisson:2011nh} or looking up the answer in
the literature \cite{bunch1978quantum,*deVega:1998ia}.  Either approach yields that \begin{multline}\label{eq:intermediate Greens} 	G(x,x') =
\frac{2M_{\star}^{2}}{H\eta' \,(M_{\star}^{2}+2H^{2})} \left[
\frac{\delta(\Delta\eta-r)}{4\pi r} +  \right. \\ \left. \frac{\theta(\Delta\eta-r)}{4\pi\eta\eta'}  +Q_{\nu}(x,x') \right], \end{multline} where
$\theta$ is the Heaviside function, $\nu^{2} =
1/4-M_{\star}^{2}/H^{2}$, 
\begin{multline}
 	Q_{\nu}(x,x') = \theta(\Delta\eta) \frac{\sqrt{\eta\eta'}}{{4\pi r}}  \int_{0}^{\infty} \!\! dk \, k
\sin kr \\ \times [ J_{\nu}(-k\eta)Y_{\nu}(-k\eta')  -
J_{\nu}(-k\eta')Y_{\nu}(-k\eta)] ,\label{eq:M formula} \end{multline} and $J_{\nu}$ and $Y_{\nu}$ are Bessel functions.  In Appendix \ref{sec:integral}, we show how it is possible, but not overly instructive, to evaluate the integral in terms of an associated Legendre function.  A more useful expression is obtained by using the asymptotic expansion of the Bessel functions for large argument and
integrating term-by-term: \begin{multline}\label{eq:Q series} 	
Q_{\nu}(x,x') = - \frac{\delta(\Delta\eta-r)}{4\pi r} - \frac{\theta(\Delta\eta-r)}{4\pi\eta\eta'} \\ 	-
\frac{(4\nu^{2}-9)\theta(\Delta\eta-r)}{32\pi\eta\eta'} \sum_{n=0}^{\infty} c_{n} \phi^{2n},
\end{multline} where $\phi = M\ell(x,x')$.  The $c_{n}$ coefficients are explicitly calculable; we find that the first few are given by
\begin{align}\nonumber 	c_{0} & = 1, & c_{1} & = \tfrac{1}{8}, \\ 	c_{2} & =
-\tfrac{4\nu^{2}-17}{768}. & c_{4} & = \tfrac{80\nu^{4}-840\nu^{2}+2381}{737280}. \end{align} Inserting (\ref{eq:Q series}) into
(\ref{eq:intermediate Greens}) yields \begin{equation}\label{eq:Greens 3} 	
G(x,x') =   \frac{M_{\star}^{2}\theta(\Delta\eta-r)}{4\pi (H\eta)(H\eta')^{2} } \sum_{n=0}^{\infty} c_{n} \phi^{2n}. \end{equation} We see that
the Green's function contains no $\delta$-functions; that is,
$G(x,x')$ is finite everywhere.  As mentioned in the Introduction, we expect $M_{\star}^{-1} \lesssim 10^{-8}\,\text{m}$ while current observations imply $H^{-1}$ is of order the size of the observable universe; i.e. we can take $M_{\star} \gg H$.   This implies that there are three regimes we can consider:
\begin{enumerate}[label={(\Roman*)}]
 \item $H|\ell| \gg 1$ (the spacetime separation between $x$ and $x'$ is greater that the cosmological horizon);
 \item $H|\ell| \ll 1 \ll M_{\star}|\ell|$ (the spacetime separation between $x$ and $x'$ is smaller than the cosmic horizon but larger than the exotic physics scale); and \label{regime 2}
 \item $M_{\star}|\ell| \ll 1$ (the spacetime separation between $x$ and $x'$ is smaller the exotic physics scale). \label{regime 3}
\end{enumerate}
We call \ref{regime 3} the ``near nullcone regime'' since it involves the smallest values of $|\ell|$. In the near nullcone region we have
 \begin{equation}\label{eq:Greens 4}
	G(x,x') =   \frac{M_{\star}^{2}\theta(\Delta\eta-r)}{4\pi}. \end{equation} This is precisely the retarded Green's function for gravitational
waves satisfying the wave equation obtained from the
linearization of (\ref{eq:field eqn}) if $G_{\alpha\beta}$ and $\Lambda$ are ignored; i.e., $(2M_{\star})^{-2} \Box^{2} \psi_{ij} = 0$
\cite{holscher}.  This confirms our heuristic argument that the behaviour of the Green's function
near the nullcone will be dominated by the Weyl-squared term in (\ref{eq:action 1}), and hence be radically different from GR.

We comment that (\ref{eq:Greens 4}) bears a striking resemblance to the Green's function of $\Box$ in $(1+1)$-dimensions.  In other words,
gravitational waves governed by (\ref{eq:action 1}) propagate as if
they were in two dimensions over short distances.  This is congruent with similar reductions in spacetime dimensionality observed in
\cite{Carlip:2009km,*Ambjorn:2005db,*Husain:2013zda}.

It is also interesting to examine (\ref{eq:Greens 3}) in regime \ref{regime 2} mentioned above.
Physically, this is the Green's function governing the propagation of gravitational waves over sub-cosmological distances larger than the exotic physics scale. Under these circumstances, the series in
(\ref{eq:Greens 3}) can be re-summed into a Bessel function $J_{1}$ of the first kind: \begin{subequations} \begin{align} 	G(x,x') & =
\frac{\theta(\Delta\eta)}{2\pi } \delta_{M_{\star}}(\sigma(x,x')), \\
\delta_{M_{\star}}(\sigma) & = \frac{\theta(-\sigma) M_{\star}J_{1} (M_{\star}\sqrt{-2\sigma})}{\sqrt{-2\sigma}}, \end{align} \end{subequations}
where $\sigma(x,x') = \frac{1}{2} \ell^{2}(x,x')$ is Synge's
world function.  It is easy to confirm that $\delta_{M_{\star}}$ is peaked about $\sigma=0$ with width $\sim 1/M_{\star}$ and that
$\int_{-\infty}^{\infty} \delta_{M_{\star}}(\sigma) d\sigma = 1$.  Hence,
$\delta_{M_{\star}}$ is a delta function sequence, and we have \begin{equation} 	\lim_{M_{\star} \rightarrow \infty} G(x,x') =
\frac{\theta(\Delta\eta)\delta(\sigma)}{2\pi } =
\frac{\delta(\Delta\eta-r)}{2\pi r}. \end{equation} This is the familiar Green's function for the wave equation $-\tfrac{1}{2} \Box \psi_{ij} =
0$; i.e., for gravitational waves in GR.  Therefore, when
$M_{\star}$ is finite but much larger than $H$, the Green's function over sub-horizon distances is a non-singular smoothed version of a
$\delta$-distribution centred on the nullcone.  Huygens' principle only holds in an approximate form: instead of sharp propagation along null
geodesics, gravitational waves will be defocussed.

It is perhaps useful to pause and discuss \emph{why} this defocussing effect occurs.  It certainly seems reasonable to assume that it rooted in the presence of the massive spin-2 mode in this type of quadratic gravity.  However, the mere existence of such a mode is not sufficient to ensure a regular Green's function on the null cone; if it were, we would expect the retarded propagator of a single massive field in $3+1$ dimensions to be non-singular, which it is not.  Actually, the essential feature is the existence two effective spin-2 degrees of freedom in the action arranged such that the total Green's function (\ref{eq:Greens 1}) is the \emph{difference} of the individual propagators.  This particular property of quadratic gravity is not expected to be unique; i.e., there ought to be other modified gravity models that exhibit similar behaviour.   Furthermore, we can conclude the spin-2 fields need not be massive to generate a finite Green's function.  This is most easily seen by the counterexample of $\Box^{2}\psi_{ij}=0$, which has a nonsingular propagator and no mass parameter.  Finally, we observe that it does not seem to be possible to capture this effect from a dispersion relation only.  For example, plane wave solutions $\psi_{ij} \propto e^{-i k_{0} t +i \mathbf{k}\cdot \mathbf{x}}$ of both $\Box \psi_{ij} = 0$ and $\Box^{2}\psi_{ij}=0$ propagate with dispersion relation $k_{0}^{2} = |\mathbf{k}|^{2}$, yet the Green's function is singular in the first case and non-singular in the second.

On the other hand, dimensionful constants like $M_{\star}$ appearing in dispersion relations will indeed dictate the timescale over which gravitational waves signal are blurred, as we now demonstrate:  Let us fix the source point
to be at the origin of 4-dimensional Minkowski coordinates $x' = (0,\mathbf{0})$, and assume the observer follows an inertial trajectory that
intersects the future lightcone of $x'$ at a point $x'' = r (1, \bm{\hat r})$.  Here, $\bm{\hat r}$ is a spatial unit vector and $r$ is the
observer's spatial distance from the source.  The observer's worldline is parametrized by
\begin{equation}
	x(\tau) = x'' + \tau \frac{(1,\mathbf{v})}{\sqrt{1-v^{2}}}, \quad v^{2} = \mathbf{v}\cdot\mathbf{v},
\end{equation}
where $\tau$ is the observer's proper time and $\mathbf{v}$ is the observer's 3-velocity in these coordinates.  Note that $\tau=0$ corresponds to
the instant when the observer crosses the light cone of $x'$.  Evaluating the Green's function on the observer's worldline yields:
\begin{equation}
	G(x(\tau),x') = \theta(\tau) \frac{M_{\star} J_{1}(M_{\star}\sqrt{\tau^{2}+2\tau r \kappa})}{\sqrt{\tau^{2}+2\tau r \kappa}},
\end{equation}
where $\kappa$ is the relativistic Doppler factor
\begin{equation}
	\kappa = \frac{1-v\cos\theta}{\sqrt{1-v^{2}}}, \quad v \cos\theta = \mathbf{v} \cdot \bm{\hat{r}};
\end{equation}
i.e., $\theta$ is the angle between the observer's 3-velocity and the line of sight from the source to $x''$.  In most astrophysical situations,
one would expect $\kappa = \mathcal{O}(1)$.  Assuming the amount of time that the observer measures the gravitational wave signal is much less
than the distance to the source (i.e.\ $0<\tau \ll \kappa r$) we have
\begin{equation}
	G(x(\tau),x') = \theta(\tau) \frac{M_{\star} J_{1}(M_{\star}\sqrt{2\tau r \kappa})}{\sqrt{2\tau r \kappa}}.
\end{equation}
This implies that the characteristic decay time $\tau_\text{d}$ of the gravitational wave signal in the observer's rest frame is
\begin{equation}
	\tau_\text{d} = \frac{1}{2\kappa M_{\star}^{2}r}.
\end{equation}
If one imposes $M_{\star} \lesssim 10^{-8}\,\text{m}$ (as suggested by laboratory tests  \cite{Will:2014kxa}) and $r$ is an astrophysically
relevant distance, then $\tau_\text{d}$ is an extremely short timescale.  Conversely, if one has a particularly small Doppler factor $\kappa
\rightarrow 0$, then it is easy to see that the characteristic decay time will be
\begin{equation}
	\tau_\text{d} = \frac{1}{M_{\star}},
\end{equation}
which will typically be much larger than in the $\kappa \sim 1$ case.  This leads us to conclude that the best case scenario for observing the
temporal broadening of the gravitational wave Green's function is when the source is highly redshifted.

It is possible get a crude estimate of $\tau_\text{d}$ for events similar to the first gravitational wave event GW150914 observed by LIGO \cite{Abbott:2016blz}, assuming that the centre of mass for such events is at rest in its home galaxy.  Consider an event at cosmological redshift $z$ at a comoving distance of $r$.  The redshift is related to the Doppler factor by $\kappa = (1+z)^{-1}$, which yields
\begin{equation}
	\frac{\tau_\text{d}}{t_{\Pl}} =  \left( \frac{1+z}{2} \right) \left( \frac{r}{l_{\Pl}} \right)^{-1} \left( \frac{M_{\star}^{-1}}{l_{\Pl}} \right)^{2},
\end{equation}
where $l_\Pl$ and $t_\Pl$ are the reduced Planck length and time, respectively.  This can be rewritten as
\begin{equation}
	\frac{\tau_\text{d}}{10^{-6}\,\text{sec}} =  \left( \frac{1+z}{2} \right) \left( \frac{r}{450\,\text{Mpc}} \right)^{-1} \left( \frac{M_{\star}^{-1}}{350\,\text{AU}} \right)^{2},
\end{equation}
GW150914 occurred at a moderate redshift of $z \approx 0.093$ and distance of $r \approx 440\,\text{Mpc}$.  If LIGO were to be able rule out any gravitational wave blurring with characteristic timescale $\gtrsim 10^{-6} \,\sec$ for this event, then we could infer that $M_{\star}^{-1} \lesssim 350 \, \text{AU}$.  While such a constraint is not competitive with laboratory tests of Newton's law, it does have the virtue of being obtained from an independent observation.  Also, we note that the LIGO detector noise curve is (relatively) high for frequencies $\gtrsim 10^{4} \,\text{Hz}$, so it might be challenging to resolve time differences of order $\tau_\text{d} \sim 10^{-6}\,\text{sec}$.  Finally, obtaining more stringent constraints on $M_{\star}$ might be possible for events that are much closer (i.e., within the Milky Way), have a high centre of mass velocity relative to the detector, or both.

\section{Summary}

We have derived the retarded Green's function for source-free gravitational waves propagating in the de Sitter vacuum of a quadratic gravity
model described by (\ref{eq:action 1}).  When the length scale of exotic physics is much smaller than the de Sitter radius, the singular Green's
function of GR is effectively smoothed-out about the nullcone.  Since the gravitational wave propagator is non-singular and has support inside
the nullcone, we conclude that Huygens' principle does not hold in this version of quadratic gravity.

The retarded Green's function can be used to write down an explicit solution of the gravitational wave initial value problem---usually called
Kirchoff's formula---in both GR \cite{Poisson:2011nh} and the current model \cite{seahra}.  In GR, one finds that gravitational waves at a given
point $x$ depend only on initial data on the point's past nullcone.  Conversely, for quadratic gravity the smoothed Green's function
(\ref{eq:Greens 3}) implies that the radiation at $x$ depends on initial data inside the past nullcone of $x$.

To grasp the significance of this fact, it is useful to imagine the same effect applied to more familiar physics:  If the Green's function of
electromagnetism were regulated as in (\ref{eq:Greens 3}), it would be impossible to see a moving object clearly; at any given time the image
detected by our eyes would be a superposition of the state of the object at multiple times.

The implication for gravitational wave observatories is clear: gravitational wave signals predicted by nonlinear simulations will blurred as
they travel from sources to detectors.  This effect will be most pronounced for highly redshifted sources where the defocussing timescale is
$M_{\star}^{-1}$.  Depending on the noise characteristics of the detector and the theoretical uncertainties in the gravitational waveforms, it
might be possible to directly constrain $M_{\star}$ independently of solar system and laboratory tests.

We note that gravitational wave sources in the early universe can plausibly have size of order  $M_{\star}^{-1}$ or smaller, and are also intrinsically highly redshifted.  Hence, the investigation of gravitational waves governed by (\ref{eq:action 1}) and produced during such epochs could be very interesting.

\appendix

\section{}\label{sec:integral}

In this Appendix, we demonstrate how (\ref{eq:M formula}) can be evaluated in terms of an associated Legendre function.  We first rewrite $Q_{\nu}(x,x')$ as
\begin{equation}
	Q_{\nu}(x,x') = -\frac{\Theta(\Delta\eta)}{4\pi r^{2}} M_{\nu} \left( - \frac{\eta}{r}, -\frac{\eta'}{r} \right),
\end{equation}
where,
\begin{multline}
	M_{\nu}(\alpha,\beta) = \sqrt{\alpha\beta} \, \text{Im} \bigg[  \int_{0}^{\infty} dz \, z \sin z \times  \\  H^{(1)}_{\nu}(\alpha z )\bar{H}^{(1)}_{\nu}(\beta z) \bigg], \quad \beta \ge \alpha >0.
\end{multline}
This can be rewritten in terms of modified Bessel functions of the second kind:
\begin{multline}
	M_{\nu}(\alpha,\beta) = \frac{4\sqrt{\alpha\beta}}{\pi^{2}} \text{Im} \bigg[  \int_{0}^{\infty} dz \, z \sin z \times  \\  K_{\nu}(-i\alpha z ) K_{\nu}(i\beta z) \bigg].
\end{multline}
The integral can be evaluated if we displace $\alpha$ and $\beta$ off the real axis
\begin{equation}
	\alpha \mapsto \alpha + i\epsilon, \quad \beta \mapsto \beta - i\epsilon,  \quad 0 < \epsilon \ll 1.
\end{equation}
Then we get \cite[eq.\ 6.692.2]{GR}
\begin{equation}
	M_{\nu}(\alpha,\beta) = - \frac{\sec \left( \pi \nu   \right)}{\alpha\beta}  \text{Im} \left[  \frac{P_{\nu-1/2}^{-1}(u+i\delta) }{\sqrt{(u+i\delta)^{2}-1}} \right],
\end{equation}
where $P_{\nu-1/2}^{-1}(u)$ is an associated Legendre function and 
\begin{align}\label{eq:u def}
	u = -1 +\frac{1-(\alpha-\beta)^{2}}{2\alpha\beta}, \quad  \delta = \frac{(\beta-\alpha)(1-u)}{\alpha\beta}\epsilon.
\end{align}
Making use of (\ref{eq:geodesic distance}), we see that
\begin{equation}\label{eq:geodesic}
	u = - \cos H\ell(x,x').
\end{equation}
It follows that $0 < \delta \ll 1$, so we obtain
\begin{equation}
	Q_{\nu}(x,x') = \frac{\Theta(\Delta\eta)\sec \left( \pi \nu   \right)}{4\pi \eta\eta'} \text{Im} \left[  \frac{P_{\nu-1/2}^{-1}(u+i0^{+}) }{\sqrt{(u+i0^{+})^{2}-1}} \right].
\end{equation}
Note that in order to actually use this expression, one needs to select the branch of $P_{\nu-1/2}^{-1}$ carefully.  For this reason, it is more straightforward to use the series expansion employed in Section \ref{sec:Green's} to obtain an explicit form of the Green's function.

\vfill

\begin{acknowledgments}

We thank Mohammad El Smaily for conversations, Shohreh Rahmati for past collaboration, and NSERC of Canada for support.  This research was
supported in part by Perimeter Institute for Theoretical Physics. Research at Perimeter Institute is supported by the Government of Canada
through Innovation, Science and Economic Development Canada and by the Province of Ontario through the Ministry of Research, Innovation and
Science.

\end{acknowledgments}

\vfill

\bibliography{GWs_YM}

\begin{thebibliography}{34}%
\makeatletter
\providecommand \@ifxundefined [1]{%
 \@ifx{#1\undefined}
}%
\providecommand \@ifnum [1]{%
 \ifnum #1\expandafter \@firstoftwo
 \else \expandafter \@secondoftwo
 \fi
}%
\providecommand \@ifx [1]{%
 \ifx #1\expandafter \@firstoftwo
 \else \expandafter \@secondoftwo
 \fi
}%
\providecommand \natexlab [1]{#1}%
\providecommand \enquote  [1]{``#1''}%
\providecommand \bibnamefont  [1]{#1}%
\providecommand \bibfnamefont [1]{#1}%
\providecommand \citenamefont [1]{#1}%
\providecommand \href@noop [0]{\@secondoftwo}%
\providecommand \href [0]{\begingroup \@sanitize@url \@href}%
\providecommand \@href[1]{\@@startlink{#1}\@@href}%
\providecommand \@@href[1]{\endgroup#1\@@endlink}%
\providecommand \@sanitize@url [0]{\catcode `\\12\catcode `\$12\catcode
  `\&12\catcode `\#12\catcode `\^12\catcode `\_12\catcode `\%12\relax}%
\providecommand \@@startlink[1]{}%
\providecommand \@@endlink[0]{}%
\providecommand \url  [0]{\begingroup\@sanitize@url \@url }%
\providecommand \@url [1]{\endgroup\@href {#1}{\urlprefix }}%
\providecommand \urlprefix  [0]{URL }%
\providecommand \Eprint [0]{\href }%
\providecommand \doibase [0]{http://dx.doi.org/}%
\providecommand \selectlanguage [0]{\@gobble}%
\providecommand \bibinfo  [0]{\@secondoftwo}%
\providecommand \bibfield  [0]{\@secondoftwo}%
\providecommand \translation [1]{[#1]}%
\providecommand \BibitemOpen [0]{}%
\providecommand \bibitemStop [0]{}%
\providecommand \bibitemNoStop [0]{.\EOS\space}%
\providecommand \EOS [0]{\spacefactor3000\relax}%
\providecommand \BibitemShut  [1]{\csname bibitem#1\endcsname}%
\let\auto@bib@innerbib\@empty
\bibitem [{\citenamefont {Abbott}\ \emph
  {et~al.}(2016{\natexlab{a}})\citenamefont {Abbott} \emph
  {et~al.}}]{Abbott:2016blz}%
  \BibitemOpen
  \bibfield  {author} {\bibinfo {author} {\bibfnamefont {B.~P.}\ \bibnamefont
  {Abbott}} \emph {et~al.} (\bibinfo {collaboration} {Virgo, LIGO
  Scientific}),\ }\href {\doibase 10.1103/PhysRevLett.116.061102} {\bibfield
  {journal} {\bibinfo  {journal} {Phys. Rev. Lett.}\ }\textbf {\bibinfo
  {volume} {116}},\ \bibinfo {pages} {061102} (\bibinfo {year}
  {2016}{\natexlab{a}})},\ \Eprint {http://arxiv.org/abs/1602.03837}
  {arXiv:1602.03837 [gr-qc]} \BibitemShut {NoStop}%
\bibitem [{\citenamefont {Yunes}\ and\ \citenamefont
  {Siemens}(2013)}]{Yunes:2013dva}%
  \BibitemOpen
  \bibfield  {author} {\bibinfo {author} {\bibfnamefont {N.}~\bibnamefont
  {Yunes}}\ and\ \bibinfo {author} {\bibfnamefont {X.}~\bibnamefont
  {Siemens}},\ }\href {\doibase 10.12942/lrr-2013-9} {\bibfield  {journal}
  {\bibinfo  {journal} {Living Rev. Rel.}\ }\textbf {\bibinfo {volume} {16}},\
  \bibinfo {pages} {9} (\bibinfo {year} {2013})},\ \Eprint
  {http://arxiv.org/abs/1304.3473} {arXiv:1304.3473 [gr-qc]} \BibitemShut
  {NoStop}%
\bibitem [{\citenamefont {Abbott}\ \emph
  {et~al.}(2016{\natexlab{b}})\citenamefont {Abbott} \emph
  {et~al.}}]{TheLIGOScientific:2016src}%
  \BibitemOpen
  \bibfield  {author} {\bibinfo {author} {\bibfnamefont {B.~P.}\ \bibnamefont
  {Abbott}} \emph {et~al.} (\bibinfo {collaboration} {Virgo, LIGO
  Scientific}),\ }\href {\doibase 10.1103/PhysRevLett.116.221101} {\bibfield
  {journal} {\bibinfo  {journal} {Phys. Rev. Lett.}\ }\textbf {\bibinfo
  {volume} {116}},\ \bibinfo {pages} {221101} (\bibinfo {year}
  {2016}{\natexlab{b}})},\ \Eprint {http://arxiv.org/abs/1602.03841}
  {arXiv:1602.03841 [gr-qc]} \BibitemShut {NoStop}%
\bibitem [{\citenamefont {Yunes}\ \emph {et~al.}(2016)\citenamefont {Yunes},
  \citenamefont {Yagi},\ and\ \citenamefont {Pretorius}}]{Yunes:2016jcc}%
  \BibitemOpen
  \bibfield  {author} {\bibinfo {author} {\bibfnamefont {N.}~\bibnamefont
  {Yunes}}, \bibinfo {author} {\bibfnamefont {K.}~\bibnamefont {Yagi}}, \ and\
  \bibinfo {author} {\bibfnamefont {F.}~\bibnamefont {Pretorius}},\ }\href
  {\doibase 10.1103/PhysRevD.94.084002} {\bibfield  {journal} {\bibinfo
  {journal} {Phys. Rev.}\ }\textbf {\bibinfo {volume} {D94}},\ \bibinfo {pages}
  {084002} (\bibinfo {year} {2016})},\ \Eprint
  {http://arxiv.org/abs/1603.08955} {arXiv:1603.08955 [gr-qc]} \BibitemShut
  {NoStop}%
\bibitem [{\citenamefont {Abbott}\ \emph
  {et~al.}(2016{\natexlab{c}})\citenamefont {Abbott} \emph
  {et~al.}}]{TheLIGOScientific:2016pea}%
  \BibitemOpen
  \bibfield  {author} {\bibinfo {author} {\bibfnamefont {B.~P.}\ \bibnamefont
  {Abbott}} \emph {et~al.} (\bibinfo {collaboration} {Virgo, LIGO
  Scientific}),\ }\href {\doibase 10.1103/PhysRevX.6.041015} {\bibfield
  {journal} {\bibinfo  {journal} {Phys. Rev.}\ }\textbf {\bibinfo {volume}
  {X6}},\ \bibinfo {pages} {041015} (\bibinfo {year} {2016}{\natexlab{c}})},\
  \Eprint {http://arxiv.org/abs/1606.04856} {arXiv:1606.04856 [gr-qc]}
  \BibitemShut {NoStop}%
\bibitem [{\citenamefont {Clifton}\ \emph {et~al.}(2012)\citenamefont
  {Clifton}, \citenamefont {Ferreira}, \citenamefont {Padilla},\ and\
  \citenamefont {Skordis}}]{Clifton:2011jh}%
  \BibitemOpen
  \bibfield  {author} {\bibinfo {author} {\bibfnamefont {T.}~\bibnamefont
  {Clifton}}, \bibinfo {author} {\bibfnamefont {P.~G.}\ \bibnamefont
  {Ferreira}}, \bibinfo {author} {\bibfnamefont {A.}~\bibnamefont {Padilla}}, \
  and\ \bibinfo {author} {\bibfnamefont {C.}~\bibnamefont {Skordis}},\ }\href
  {\doibase 10.1016/j.physrep.2012.01.001} {\bibfield  {journal} {\bibinfo
  {journal} {Phys. Rept.}\ }\textbf {\bibinfo {volume} {513}},\ \bibinfo
  {pages} {1} (\bibinfo {year} {2012})},\ \Eprint
  {http://arxiv.org/abs/1106.2476} {arXiv:1106.2476 [astro-ph.CO]} \BibitemShut
  {NoStop}%
\bibitem [{\citenamefont {Berti}\ \emph {et~al.}(2015)\citenamefont {Berti}
  \emph {et~al.}}]{Berti:2015itd}%
  \BibitemOpen
  \bibfield  {author} {\bibinfo {author} {\bibfnamefont {E.}~\bibnamefont
  {Berti}} \emph {et~al.},\ }\href {\doibase 10.1088/0264-9381/32/24/243001}
  {\bibfield  {journal} {\bibinfo  {journal} {Class. Quant. Grav.}\ }\textbf
  {\bibinfo {volume} {32}},\ \bibinfo {pages} {243001} (\bibinfo {year}
  {2015})},\ \Eprint {http://arxiv.org/abs/1501.07274} {arXiv:1501.07274
  [gr-qc]} \BibitemShut {NoStop}%
\bibitem [{\citenamefont {Sotiriou}\ and\ \citenamefont
  {Faraoni}(2010)}]{Sotiriou:2008rp}%
  \BibitemOpen
  \bibfield  {author} {\bibinfo {author} {\bibfnamefont {T.~P.}\ \bibnamefont
  {Sotiriou}}\ and\ \bibinfo {author} {\bibfnamefont {V.}~\bibnamefont
  {Faraoni}},\ }\href {\doibase 10.1103/RevModPhys.82.451} {\bibfield
  {journal} {\bibinfo  {journal} {Rev. Mod. Phys.}\ }\textbf {\bibinfo {volume}
  {82}},\ \bibinfo {pages} {451} (\bibinfo {year} {2010})},\ \Eprint
  {http://arxiv.org/abs/0805.1726} {arXiv:0805.1726 [gr-qc]} \BibitemShut
  {NoStop}%
\bibitem [{\citenamefont {Stelle}(1977)}]{Stelle:1976gc}%
  \BibitemOpen
  \bibfield  {author} {\bibinfo {author} {\bibfnamefont {K.~S.}\ \bibnamefont
  {Stelle}},\ }\href {\doibase 10.1103/PhysRevD.16.953} {\bibfield  {journal}
  {\bibinfo  {journal} {Phys. Rev.}\ }\textbf {\bibinfo {volume} {D16}},\
  \bibinfo {pages} {953} (\bibinfo {year} {1977})}\BibitemShut {NoStop}%
\bibitem [{\citenamefont {Stelle}(1978)}]{Stelle:1977ry}%
  \BibitemOpen
  \bibfield  {author} {\bibinfo {author} {\bibfnamefont {K.~S.}\ \bibnamefont
  {Stelle}},\ }\href {\doibase 10.1007/BF00760427} {\bibfield  {journal}
  {\bibinfo  {journal} {Gen. Rel. Grav.}\ }\textbf {\bibinfo {volume} {9}},\
  \bibinfo {pages} {353} (\bibinfo {year} {1978})}\BibitemShut {NoStop}%
\bibitem [{\citenamefont {Boulware}\ \emph {et~al.}(1983)\citenamefont
  {Boulware}, \citenamefont {Horowitz},\ and\ \citenamefont
  {Strominger}}]{Boulware:1983td}%
  \BibitemOpen
  \bibfield  {author} {\bibinfo {author} {\bibfnamefont {D.~G.}\ \bibnamefont
  {Boulware}}, \bibinfo {author} {\bibfnamefont {G.~T.}\ \bibnamefont
  {Horowitz}}, \ and\ \bibinfo {author} {\bibfnamefont {A.}~\bibnamefont
  {Strominger}},\ }\href {\doibase 10.1103/PhysRevLett.50.1726} {\bibfield
  {journal} {\bibinfo  {journal} {Phys. Rev. Lett.}\ }\textbf {\bibinfo
  {volume} {50}},\ \bibinfo {pages} {1726} (\bibinfo {year}
  {1983})}\BibitemShut {NoStop}%
\bibitem [{\citenamefont {David}\ and\ \citenamefont
  {Strominger}(1984)}]{David:1984uv}%
  \BibitemOpen
  \bibfield  {author} {\bibinfo {author} {\bibfnamefont {F.}~\bibnamefont
  {David}}\ and\ \bibinfo {author} {\bibfnamefont {A.}~\bibnamefont
  {Strominger}},\ }\href {\doibase 10.1016/0370-2693(84)90817-7} {\bibfield
  {journal} {\bibinfo  {journal} {Phys. Lett.}\ }\textbf {\bibinfo {volume}
  {B143}},\ \bibinfo {pages} {125} (\bibinfo {year} {1984})}\BibitemShut
  {NoStop}%
\bibitem [{\citenamefont {Horowitz}(1985)}]{Horowitz:1984wv}%
  \BibitemOpen
  \bibfield  {author} {\bibinfo {author} {\bibfnamefont {G.~T.}\ \bibnamefont
  {Horowitz}},\ }\href {\doibase 10.1103/PhysRevD.31.1169} {\bibfield
  {journal} {\bibinfo  {journal} {Phys. Rev.}\ }\textbf {\bibinfo {volume}
  {D31}},\ \bibinfo {pages} {1169} (\bibinfo {year} {1985})}\BibitemShut
  {NoStop}%
\bibitem [{\citenamefont {Deser}\ and\ \citenamefont
  {Tekin}(2003)}]{Deser:2003up}%
  \BibitemOpen
  \bibfield  {author} {\bibinfo {author} {\bibfnamefont {S.}~\bibnamefont
  {Deser}}\ and\ \bibinfo {author} {\bibfnamefont {B.}~\bibnamefont {Tekin}},\
  }\href {\doibase 10.1088/0264-9381/20/22/011} {\bibfield  {journal} {\bibinfo
   {journal} {Class. Quant. Grav.}\ }\textbf {\bibinfo {volume} {20}},\
  \bibinfo {pages} {4877} (\bibinfo {year} {2003})},\ \Eprint
  {http://arxiv.org/abs/gr-qc/0306114} {arXiv:gr-qc/0306114 [gr-qc]}
  \BibitemShut {NoStop}%
\bibitem [{\citenamefont {Deser}\ and\ \citenamefont
  {Tekin}(2007)}]{Deser:2007vs}%
  \BibitemOpen
  \bibfield  {author} {\bibinfo {author} {\bibfnamefont {S.}~\bibnamefont
  {Deser}}\ and\ \bibinfo {author} {\bibfnamefont {B.}~\bibnamefont {Tekin}},\
  }\href {\doibase 10.1103/PhysRevD.75.084032} {\bibfield  {journal} {\bibinfo
  {journal} {Phys. Rev.}\ }\textbf {\bibinfo {volume} {D75}},\ \bibinfo {pages}
  {084032} (\bibinfo {year} {2007})},\ \Eprint
  {http://arxiv.org/abs/gr-qc/0701140} {arXiv:gr-qc/0701140 [gr-qc]}
  \BibitemShut {NoStop}%
\bibitem [{\citenamefont {'t~Hooft}(2011)}]{tHooft:2011aa}%
  \BibitemOpen
  \bibfield  {author} {\bibinfo {author} {\bibfnamefont {G.}~\bibnamefont
  {'t~Hooft}},\ }\href {\doibase 10.1007/s10701-011-9586-8} {\bibfield
  {journal} {\bibinfo  {journal} {Found. Phys.}\ }\textbf {\bibinfo {volume}
  {41}},\ \bibinfo {pages} {1829} (\bibinfo {year} {2011})},\ \Eprint
  {http://arxiv.org/abs/1104.4543} {arXiv:1104.4543 [gr-qc]} \BibitemShut
  {NoStop}%
\bibitem [{\citenamefont {Maldacena}(2011)}]{Maldacena:2011mk}%
  \BibitemOpen
  \bibfield  {author} {\bibinfo {author} {\bibfnamefont {J.}~\bibnamefont
  {Maldacena}},\ }\href@noop {} {\  (\bibinfo {year} {2011})},\ \Eprint
  {http://arxiv.org/abs/1105.5632} {arXiv:1105.5632 [hep-th]} \BibitemShut
  {NoStop}%
\bibitem [{\citenamefont {Capozziello}\ and\ \citenamefont
  {Stabile}(2015)}]{Capozziello:2015nga}%
  \BibitemOpen
  \bibfield  {author} {\bibinfo {author} {\bibfnamefont {S.}~\bibnamefont
  {Capozziello}}\ and\ \bibinfo {author} {\bibfnamefont {A.}~\bibnamefont
  {Stabile}},\ }\href {\doibase 10.1007/s10509-015-2425-1} {\bibfield
  {journal} {\bibinfo  {journal} {Astrophys. Space Sci.}\ }\textbf {\bibinfo
  {volume} {358}},\ \bibinfo {pages} {27} (\bibinfo {year} {2015})}\BibitemShut
  {NoStop}%
\bibitem [{\citenamefont {Hinterbichler}\ and\ \citenamefont
  {Saravani}(2016)}]{Hinterbichler:2015soa}%
  \BibitemOpen
  \bibfield  {author} {\bibinfo {author} {\bibfnamefont {K.}~\bibnamefont
  {Hinterbichler}}\ and\ \bibinfo {author} {\bibfnamefont {M.}~\bibnamefont
  {Saravani}},\ }\href {\doibase 10.1103/PhysRevD.93.065006} {\bibfield
  {journal} {\bibinfo  {journal} {Phys. Rev.}\ }\textbf {\bibinfo {volume}
  {D93}},\ \bibinfo {pages} {065006} (\bibinfo {year} {2016})},\ \Eprint
  {http://arxiv.org/abs/1508.02401} {arXiv:1508.02401 [hep-th]} \BibitemShut
  {NoStop}%
\bibitem [{\citenamefont {Gegenberg}\ \emph {et~al.}(2016)\citenamefont
  {Gegenberg}, \citenamefont {Rahmati},\ and\ \citenamefont
  {Seahra}}]{Gegenberg:2015gma}%
  \BibitemOpen
  \bibfield  {author} {\bibinfo {author} {\bibfnamefont {J.}~\bibnamefont
  {Gegenberg}}, \bibinfo {author} {\bibfnamefont {S.}~\bibnamefont {Rahmati}},
  \ and\ \bibinfo {author} {\bibfnamefont {S.~S.}\ \bibnamefont {Seahra}},\
  }\href {\doibase 10.1103/PhysRevD.93.064025} {\bibfield  {journal} {\bibinfo
  {journal} {Phys. Rev.}\ }\textbf {\bibinfo {volume} {D93}},\ \bibinfo {pages}
  {064025} (\bibinfo {year} {2016})},\ \Eprint
  {http://arxiv.org/abs/1505.06058} {arXiv:1505.06058 [gr-qc]} \BibitemShut
  {NoStop}%
\bibitem [{\citenamefont {Alvarez-Gaume}\ \emph {et~al.}(2016)\citenamefont
  {Alvarez-Gaume}, \citenamefont {Kehagias}, \citenamefont {Kounnas},
  \citenamefont {L{\"u}st},\ and\ \citenamefont
  {Riotto}}]{Alvarez-Gaume:2015rwa}%
  \BibitemOpen
  \bibfield  {author} {\bibinfo {author} {\bibfnamefont {L.}~\bibnamefont
  {Alvarez-Gaume}}, \bibinfo {author} {\bibfnamefont {A.}~\bibnamefont
  {Kehagias}}, \bibinfo {author} {\bibfnamefont {C.}~\bibnamefont {Kounnas}},
  \bibinfo {author} {\bibfnamefont {D.}~\bibnamefont {L{\"u}st}}, \ and\
  \bibinfo {author} {\bibfnamefont {A.}~\bibnamefont {Riotto}},\ }\href
  {\doibase 10.1002/prop.201500100} {\bibfield  {journal} {\bibinfo  {journal}
  {Fortsch. Phys.}\ }\textbf {\bibinfo {volume} {64}},\ \bibinfo {pages} {176}
  (\bibinfo {year} {2016})},\ \Eprint {http://arxiv.org/abs/1505.07657}
  {arXiv:1505.07657 [hep-th]} \BibitemShut {NoStop}%
\bibitem [{\citenamefont {Berry}\ and\ \citenamefont
  {Gair}(2011)}]{Berry:2011pb}%
  \BibitemOpen
  \bibfield  {author} {\bibinfo {author} {\bibfnamefont {C.~P.~L.}\
  \bibnamefont {Berry}}\ and\ \bibinfo {author} {\bibfnamefont {J.~R.}\
  \bibnamefont {Gair}},\ }\href {\doibase 10.1103/PhysRevD.85.089906,
  10.1103/PhysRevD.83.104022} {\bibfield  {journal} {\bibinfo  {journal} {Phys.
  Rev.}\ }\textbf {\bibinfo {volume} {D83}},\ \bibinfo {pages} {104022}
  (\bibinfo {year} {2011})},\ \bibinfo {note} {[Erratum: Phys.
  Rev.D85,089906(2012)]},\ \Eprint {http://arxiv.org/abs/1104.0819}
  {arXiv:1104.0819 [gr-qc]} \BibitemShut {NoStop}%
\bibitem [{\citenamefont {'t~Hooft}(2014)}]{Hooft:2014daa}%
  \BibitemOpen
  \bibfield  {author} {\bibinfo {author} {\bibfnamefont {G.}~\bibnamefont
  {'t~Hooft}},\ }\href@noop {} {\  (\bibinfo {year} {2014})},\ \Eprint
  {http://arxiv.org/abs/1410.6675} {arXiv:1410.6675 [gr-qc]} \BibitemShut
  {NoStop}%
\bibitem [{\citenamefont {Will}(2014)}]{Will:2014kxa}%
  \BibitemOpen
  \bibfield  {author} {\bibinfo {author} {\bibfnamefont {C.~M.}\ \bibnamefont
  {Will}},\ }\href {\doibase 10.12942/lrr-2014-4} {\bibfield  {journal}
  {\bibinfo  {journal} {Living Rev. Rel.}\ }\textbf {\bibinfo {volume} {17}},\
  \bibinfo {pages} {4} (\bibinfo {year} {2014})},\ \Eprint
  {http://arxiv.org/abs/1403.7377} {arXiv:1403.7377 [gr-qc]} \BibitemShut
  {NoStop}%
\bibitem [{\citenamefont {Tsamis}\ and\ \citenamefont
  {Woodard}(1994)}]{Tsamis:1992xa}%
  \BibitemOpen
  \bibfield  {author} {\bibinfo {author} {\bibfnamefont {N.~C.}\ \bibnamefont
  {Tsamis}}\ and\ \bibinfo {author} {\bibfnamefont {R.~P.}\ \bibnamefont
  {Woodard}},\ }\href {\doibase 10.1007/BF02102015} {\bibfield  {journal}
  {\bibinfo  {journal} {Commun. Math. Phys.}\ }\textbf {\bibinfo {volume}
  {162}},\ \bibinfo {pages} {217} (\bibinfo {year} {1994})}\BibitemShut
  {NoStop}%
\bibitem [{\citenamefont {Poisson}\ \emph {et~al.}(2011)\citenamefont
  {Poisson}, \citenamefont {Pound},\ and\ \citenamefont
  {Vega}}]{Poisson:2011nh}%
  \BibitemOpen
  \bibfield  {author} {\bibinfo {author} {\bibfnamefont {E.}~\bibnamefont
  {Poisson}}, \bibinfo {author} {\bibfnamefont {A.}~\bibnamefont {Pound}}, \
  and\ \bibinfo {author} {\bibfnamefont {I.}~\bibnamefont {Vega}},\ }\href
  {\doibase 10.12942/lrr-2011-7} {\bibfield  {journal} {\bibinfo  {journal}
  {Living Rev. Rel.}\ }\textbf {\bibinfo {volume} {14}},\ \bibinfo {pages} {7}
  (\bibinfo {year} {2011})},\ \Eprint {http://arxiv.org/abs/1102.0529}
  {arXiv:1102.0529 [gr-qc]} \BibitemShut {NoStop}%
\bibitem [{\citenamefont {Bunch}\ and\ \citenamefont
  {Davies}(1978)}]{bunch1978quantum}%
  \BibitemOpen
  \bibfield  {author} {\bibinfo {author} {\bibfnamefont {T.~S.}\ \bibnamefont
  {Bunch}}\ and\ \bibinfo {author} {\bibfnamefont {P.~C.}\ \bibnamefont
  {Davies}},\ }in\ \href@noop {} {\emph {\bibinfo {booktitle} {Proceedings of
  the Royal Society of London A: Mathematical, Physical and Engineering
  Sciences}}},\ Vol.\ \bibinfo {volume} {360}\ (\bibinfo {organization} {The
  Royal Society},\ \bibinfo {year} {1978})\ pp.\ \bibinfo {pages}
  {117--134}\BibitemShut {NoStop}%
\bibitem [{\citenamefont {de~Vega}\ \emph {et~al.}(1999)\citenamefont
  {de~Vega}, \citenamefont {Ramirez},\ and\ \citenamefont
  {Sanchez}}]{deVega:1998ia}%
  \BibitemOpen
  \bibfield  {author} {\bibinfo {author} {\bibfnamefont {H.~J.}\ \bibnamefont
  {de~Vega}}, \bibinfo {author} {\bibfnamefont {J.}~\bibnamefont {Ramirez}}, \
  and\ \bibinfo {author} {\bibfnamefont {N.~G.}\ \bibnamefont {Sanchez}},\
  }\href {\doibase 10.1103/PhysRevD.60.044007} {\bibfield  {journal} {\bibinfo
  {journal} {Phys. Rev.}\ }\textbf {\bibinfo {volume} {D60}},\ \bibinfo {pages}
  {044007} (\bibinfo {year} {1999})},\ \Eprint
  {http://arxiv.org/abs/astro-ph/9812465} {arXiv:astro-ph/9812465 [astro-ph]}
  \BibitemShut {NoStop}%
\bibitem [{\citenamefont {H\"olscher}(2015)}]{holscher}%
  \BibitemOpen
  \bibfield  {author} {\bibinfo {author} {\bibfnamefont {P.}~\bibnamefont
  {H\"olscher}},\ }\emph {\bibinfo {title} {Conformal Gravity}},\ \href@noop {}
  {Master's thesis},\ \bibinfo  {school} {University of Bielefeld} (\bibinfo
  {year} {2015})\BibitemShut {NoStop}%
\bibitem [{\citenamefont {Carlip}(2009)}]{Carlip:2009km}%
  \BibitemOpen
  \bibfield  {author} {\bibinfo {author} {\bibfnamefont {S.}~\bibnamefont
  {Carlip}},\ }in\ \href
  {https://inspirehep.net/record/867166/files/arXiv:1009.1136.pdf} {\emph
  {\bibinfo {booktitle} {{Proceedings, Foundations of Space and Time:
  Reflections on Quantum Gravity: Cape Town, South Africa}}}}\ (\bibinfo {year}
  {2009})\ pp.\ \bibinfo {pages} {69--84},\ \Eprint
  {http://arxiv.org/abs/1009.1136} {arXiv:1009.1136 [gr-qc]} \BibitemShut
  {NoStop}%
\bibitem [{\citenamefont {Ambjorn}\ \emph {et~al.}(2005)\citenamefont
  {Ambjorn}, \citenamefont {Jurkiewicz},\ and\ \citenamefont
  {Loll}}]{Ambjorn:2005db}%
  \BibitemOpen
  \bibfield  {author} {\bibinfo {author} {\bibfnamefont {J.}~\bibnamefont
  {Ambjorn}}, \bibinfo {author} {\bibfnamefont {J.}~\bibnamefont {Jurkiewicz}},
  \ and\ \bibinfo {author} {\bibfnamefont {R.}~\bibnamefont {Loll}},\ }\href
  {\doibase 10.1103/PhysRevLett.95.171301} {\bibfield  {journal} {\bibinfo
  {journal} {Phys. Rev. Lett.}\ }\textbf {\bibinfo {volume} {95}},\ \bibinfo
  {pages} {171301} (\bibinfo {year} {2005})},\ \Eprint
  {http://arxiv.org/abs/hep-th/0505113} {arXiv:hep-th/0505113 [hep-th]}
  \BibitemShut {NoStop}%
\bibitem [{\citenamefont {Husain}\ \emph {et~al.}(2013)\citenamefont {Husain},
  \citenamefont {Seahra},\ and\ \citenamefont {Webster}}]{Husain:2013zda}%
  \BibitemOpen
  \bibfield  {author} {\bibinfo {author} {\bibfnamefont {V.}~\bibnamefont
  {Husain}}, \bibinfo {author} {\bibfnamefont {S.~S.}\ \bibnamefont {Seahra}},
  \ and\ \bibinfo {author} {\bibfnamefont {E.~J.}\ \bibnamefont {Webster}},\
  }\href {\doibase 10.1103/PhysRevD.88.024014} {\bibfield  {journal} {\bibinfo
  {journal} {Phys. Rev.}\ }\textbf {\bibinfo {volume} {D88}},\ \bibinfo {pages}
  {024014} (\bibinfo {year} {2013})},\ \Eprint {http://arxiv.org/abs/1305.2814}
  {arXiv:1305.2814 [hep-th]} \BibitemShut {NoStop}%
\bibitem [{\citenamefont {Seahra}(2017)}]{seahra}%
  \BibitemOpen
  \bibfield  {author} {\bibinfo {author} {\bibfnamefont {S.~S.}\ \bibnamefont
  {Seahra}},\ }\href@noop {} {\enquote {\bibinfo {title} {On the well-posedness
  of the initial value problem for d'{A}lembertian squared wave equations},}\ }
  (\bibinfo {year} {2017}),\ \bibinfo {note} {in preparation}\BibitemShut
  {NoStop}%
\bibitem [{\citenamefont {Gradshteyn}\ and\ \citenamefont {Ryzhik}(1980)}]{GR}%
  \BibitemOpen
  \bibfield  {author} {\bibinfo {author} {\bibfnamefont {I.~S.}\ \bibnamefont
  {Gradshteyn}}\ and\ \bibinfo {author} {\bibfnamefont {I.~M.}\ \bibnamefont
  {Ryzhik}},\ }\href@noop {} {\emph {\bibinfo {title} {Table of Integrals,
  Series, and Products}}},\ \bibinfo {edition} {5th}\ ed.,\ edited by\ \bibinfo
  {editor} {\bibfnamefont {A.}~\bibnamefont {Jeffery}}\ (\bibinfo  {publisher}
  {Academic Press},\ \bibinfo {address} {London},\ \bibinfo {year}
  {1980})\BibitemShut {NoStop}%
\end{thebibliography}%

\end{document}